\documentclass[aps,prl,twocolumn,superscriptaddress]{revtex4-1}
\usepackage{blindtext}
\usepackage{graphicx}% Include figure files
\usepackage{dcolumn}% Align table columns on decimal point
\usepackage{bm}
\usepackage[english]{babel}
\usepackage{color}
\usepackage{xcolor}
\usepackage{float}
\usepackage[utf8]{inputenc}
\usepackage[T1]{fontenc}
\usepackage{soul}
\usepackage[normalem]{ulem}

\usepackage{amsmath}

\makeatother
\begin{document}

% \title{Protocol for the Construction of a Polydisperse Ideal Jammed Packing in two Dimensions}
% \title{Constructing the Ideal Glass and Ideal Jammed Disk Packing in Two Dimensions}
\title{The Ideal Glass and the Ideal Disk Packing in Two Dimensions}

\author{Viola M. Bolton-Lum}
\affiliation{Department of Physics and Materials Science Institute,
University of Oregon, Eugene, Oregon 97403, USA.}
\author{R. Cameron Dennis}
\affiliation{Department of Physics and Astronomy, University of Pennsylvania, 209 South 33rd Street, Philadelphia,
Pennsylvania 19104, USA}
\affiliation{Department of Physics and Soft Matter Program, Syracuse University, Syracuse, New York 13244, USA}
\author{Peter K. Morse}
\affiliation{Department of Physics, Seton Hall University, South Orange, NJ 07079}
\affiliation{Department of Chemistry, Department of Physics and Princeton Institute of Materials, Princeton University, Princeton, New Jersey 08544, USA}
\author{Eric I. Corwin}
\affiliation{Department of Physics and Materials Science Institute,
University of Oregon, Eugene, Oregon 97403, USA.}
\date{\today}

\begin{abstract}
The ideal glass, a disordered system of particles with zero configurational entropy, cannot be realized through thermal processes.  Nevertheless, we present a method for constructing ideal jammed packings of soft spheres, and thus the zero temperature ideal glass, in two dimensions. In line with the predicted properties, these critically jammed packings have high bulk and shear moduli as well as an anomalously high density. While the absence of pressure scaling in the shear moduli of crystalline materials is often attributed to the ordered nature of the particles, we show for the first time that disordered ideal packings also have this feature. We also find that the density of states avoids the low frequency power law scaling famously found in most amorphous materials, these configurations display hyperuniformity, and they melt at unusually high temperatures as compared to conventional packings. In addition to resolving a long-standing mystery, this methodology represents a valuable shortcut in the generation of well-equilibrated glassy systems. The creation of such an ideal packing makes possible a complete exploration and explanation of two dimensional jammed and glassy systems.
\end{abstract}

\maketitle

\textit{Introduction} --
A defect free crystal is the apex of order~\cite{planck_vorlesungen_1905}. By contrast, when a liquid is cooled rapidly, it forms a glass, an amorphous phase of matter which is the epitome of disorder\cite{ohern_jamming_2003,turnbull_under_1969,ediger_supercooled_1996, debenedetti_supercooled_2001,charbonneau_glass_2017}. Nearly all~\cite{smallenburg_liquids_2013} liquids have a larger entropy than crystalline solids but lose entropy much faster as temperature is decreased~\cite{kauzmann_nature_1948,berthier_theoretical_2011,charbonneau_hopping_2014}. In 1948, Kauzmann recognized that there must exist a temperature at which the entropy of the liquid will cross the entropy of the crystal, but dismissed as paradoxical the possibility of such an ``ideal glass''~\cite{kauzmann_nature_1948}. How could there be a liquid state which is both amorphous and highly ordered~\cite{royall_race_2018,berthier_theoretical_2011,cavagna_supercooled_2009}? Here, we show a non-equilibrium mechanism for creating a two-dimensional zero configurational entropy jammed packing of polydisperse disks at zero temperature, which we term an ideal jammed packing. Such a packing represents the zero temperature limit of an ideal thermal glass, which is unreachable by physical processes due to a divergence in relaxation time~\cite{berthier_configurational_2017,royall_race_2018}. This mechanism exploits transient degrees of freedom~\cite{gazzillo_equation_1989, ediger_perspective_2017, ikeda_mean_2017, ninarello_models_2017, szamel_theory_2018, rodriguez-tinoco_ultrastable_2022, kim_structural_2022, hagh_transient_2022,kim_exceptionally_2024,kapteijns_fast_2019,brito_theory_2018,wang_hyperuniform_2025} and leads to zero configurational entropy at jamming, which is equivalent to our proposed definition of an ideal jammed packing as 1) having a fully triangulated contact network, 2) lacking long range crystalline orientational and translational order, 3) being mechanically ultrastable, 4) being hyperuniform, and 5) having an anomalously high melting temperature equal to the Kauzmann temperature and an anomalously low melting density. We demonstrate that this ``ideal'' state has all of the mechanical and thermal properties of a crystal, whilst entirely devoid of crystalline order. This work not only resolves the Kauzmann paradox but also demonstrates the existence of a thermodynamic glass phase for this two-dimensional model glass forming system. The creation of such an ideal glass or packing makes possible a complete exploration and explanation of two dimensional jammed and glassy systems.

\textit{Packing Protocols} -- 
%\section{Packing Protocols}
A system of $N$ disks in 2-dimensions has $2N$ translational degrees of freedom. At jamming the system is isostatic, which means that every degree of freedom is constrained and thus the system must have 2N contacts between disks~\cite{maxwell_l_1864,hagh_disordered_2018}.  Since every contact is shared by two disks, each disk must then have an average coordination number of 4.  A triangulated packing, by contrast, must have an average coordination number of 6 and thus be maximally hyperstatic~\cite{stephenson_introduction_2005}.  We achieve this by adding an additional degree of freedom to every disk in the form of a mutable radius.

The following protocol employs radii degrees of freedom to construct fully coordinated ideal amorphous packings in two dimensions while approximately preserving an input distribution of radii. This protocol builds on the use of transient degrees of freedom~\cite{hagh_transient_2022} and conformal circle packing~\cite{stephenson_introduction_2005}. 

\textit{Triangulated packings} -- $N$ disks are placed randomly into a square simulation box with side length 1 and periodic boundary conditions.  Disks are assigned radii drawn from a log-normal distribution with 20\% polydispersity~\cite{ninarello_models_2017}, chosen to avoid crystallization~\cite{tong_crystals_2015}.  Radii are scaled to an initial packing fraction $\varphi = 0.915$, chosen to result in highly coordinated (i.e. hyperstatic) packings~\cite{ohern_random_2002,ohern_jamming_2003}. The chosen value of $\varphi$ leads to highly overjammed packings, which may be produced more quickly than critically jammed packings and are thus computationally convenient. Unjammed or critically jammed packings may also be used, producing qualitatively similar results. Particles interact through the soft sphere harmonic interaction potential
\begin{equation}
    U \equiv \sum_{\left<i j\right>}\frac{1}{2}\left(1-\frac{r_{ij}}{\sigma_{ij}}\right)^2 \Theta\left(1-\frac{r_{ij}}{\sigma_{ij}}\right), 
\end{equation}
where the sum is over pairs of indices $\left<ij\right>$, $\sigma_{ij}$ is the sum of the $i$th and $j$th radii and $r_{ij}$ is the distance between centers. We use PyCudaPacking~\cite{charbonneau_jamming_2015,morse_geometric_2014}, a GPU based software package with an implementation of the FIRE algorithm~\cite{bitzek_structural_2006}, to adjust the positions and radii of the disks to find the inherent state structure, a local minimum of energy. Similarly to Hagh \textit{et al.}~\cite{hagh_transient_2022}, the radii degrees of freedom are subject to constraints on moments of the radius distribution, in order to avoid falling into various trivial global minima. The result of such a minimization is nearly triangulated, however these additional constraints necessarily introduce at least an equal number of ``defects'' into our system in the form of missing contacts.

\begin{figure}[h]
\centering
\includegraphics[trim={55 0 55 0}, clip, width=.49\linewidth]{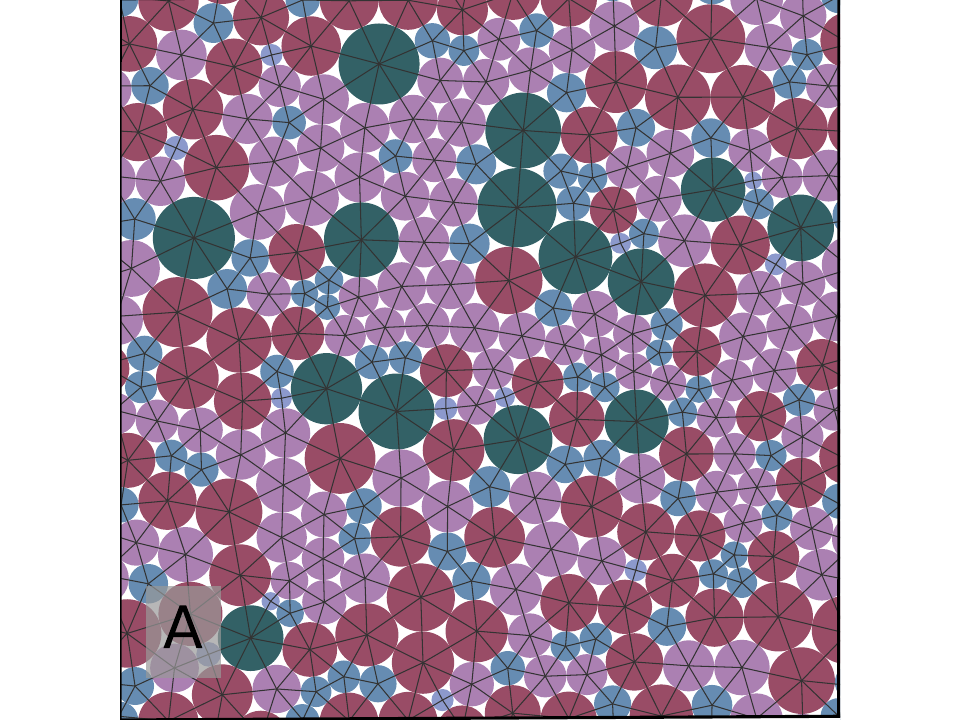}    
\includegraphics[trim={55 0 55 0}, clip, width=.49\linewidth]{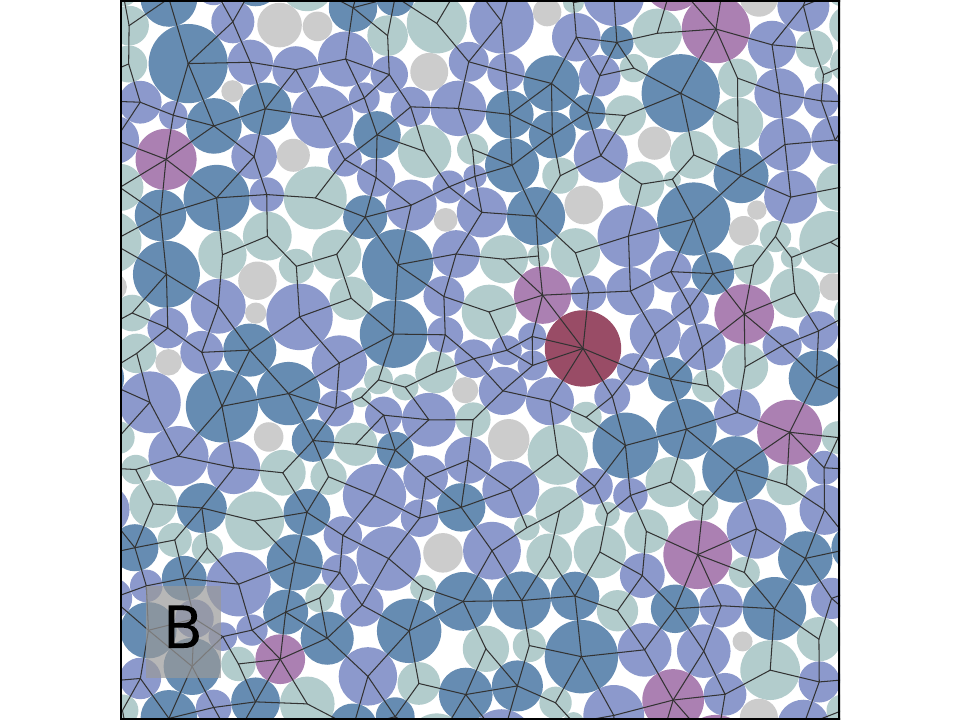}
\vspace{1pt}
\includegraphics[trim={12 16 8 2}, clip, width=.49\linewidth]{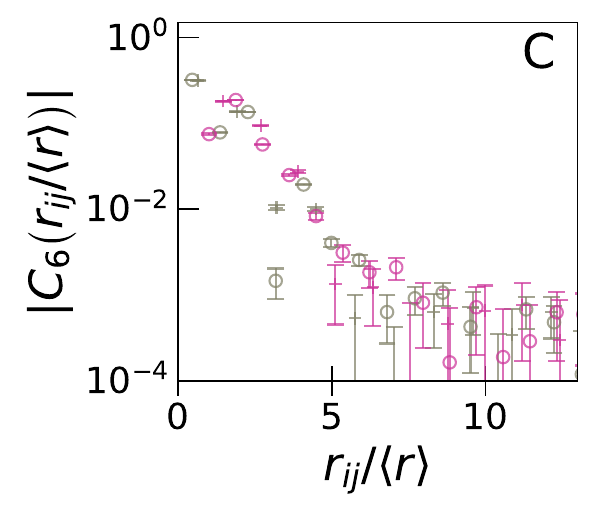}
\includegraphics[trim={12 16 8 2}, clip, width=.49\linewidth]{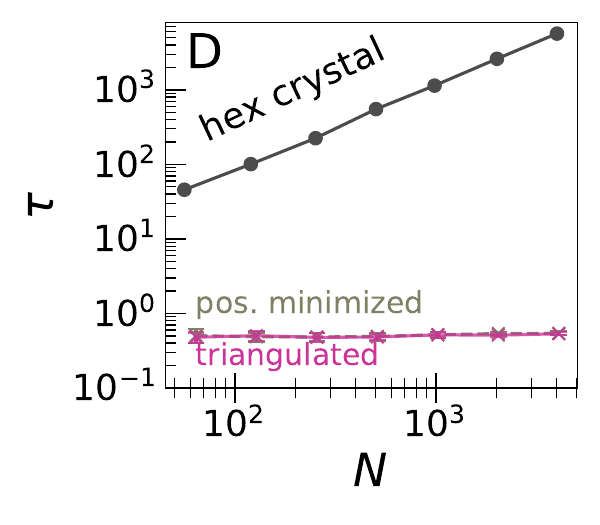}
    \caption{a) Triangulated packing, and b) conventionally jammed packing with $N=256$ and an identical set of radii. The contact network is overlaid in black, while face coloring of disks corresponds to the number of contacts contacting neighbors (grey for zero, light green for three, periwinkle for four, blue for five, purple for six, red for seven and teal for eight. c) The orientational correlation function, $C_6$, for triangulated packings (magenta and conventionally jammed packings (tan) of $N=2048$ (+'s) and $4096$ (o's). d) The finite size scaling for the translational order parameter, $\tau$, for triangulated packings (magenta), conventionally jammed packings (tan), and the hexagonal crystal (black).} %e) The probability distribution of disk diameters, $P(\sigma)/\langle \sigma \rangle)$ averaged over 10 packings of 4096 disks.  The starting distribution is at 20\% polydispersity (gray dashed line), the distribution after radii-minimization is plotted as gray x's, the distribution after triangulation is plotted as magenta triangles, and for comparison a 26\% polydispersity distribution is shown in black.}

    \label{fig:PackingWithPhi6}
\end{figure}

In order to produce fully triangulated packings, we compute the radical Delaunay triangulation for the radii-minimized configurations. To find a fully triangulated hard sphere packing, we define a constraint for each edge in the radical Delaunay network requiring the distance between the centerpoints of contacting particles to be equal to the sum of their radii:
\begin{align}
    g_{\langle ij\rangle}\equiv \sum_{\alpha=1}^2\left(x_j^{\alpha}-x_i^{\alpha}+\sum_{\beta=1}^2\Lambda_{\beta}^{\alpha}z_{ij}^{\beta}\right)^2 - \left(r_i + r_j\right)^2
\end{align}
where $x_i^{\alpha}$ is the position of particle $i$ in dimension $\alpha,$ $r_i$ is the radius of particle $i$, $\Lambda$ is the lattice vector matrix, and $z^{\beta}_{ij}$ is the $\beta^{\textrm{th}}$ component of the vector of integer lattice coordinates indicating which periodic image of particle $i$ will be closest to the central image of particle $j$. Because the triangulated network is a maximal planar graph, the circle packing theorem guarantees the existence of a unique configuration which satisfies these constraints~\cite{stephenson_introduction_2005}. We write all of the degrees of freedom for our system as a single vector, $\vec{c},$ and place an artificial spring-like attraction to the radii minimized system, $\vec{c}_0,$ to get the following Lagrangian:
\begin{align}
    \mathcal{L}\equiv \left|\vec{c}-\vec{c}_{0}\right|^2+\sum_{\langle ij\rangle}\lambda_{\langle ij \rangle}g_{\langle ij \rangle}
\end{align}
where $\lambda$ are the Lagrange multipliers. The total number of degrees of freedom in the system will be 2$N$ (positions) + $N$ (radii) + 4 (lattice vectors).  However, from this we subtract the 2 (bulk translations) + 1 (bulk rotation) + 1 (overall scale).  This leaves us with 3$N$ degrees of freedom and 3$N$ constraints for a triangulated graph. This method produces critically jammed packings with no gaps between contacting particles and overlaps that are $\mathcal{O}(10^{-30})$. One may alternatively employ CirclePack{}~\cite{stephenson_introduction_2005} to produce a new disk packing from the radical Delaunay triangulation, if this level of precision is not required. 

Our protocols yield packings at jamming fraction $\varphi \simeq 0.910$, the densest such 2d amorphous packings of which we are aware.  This density is meaningfully larger than that of the size-segregated hexagonal lattice of disks, $\varphi_\textrm{hex} = \pi/\sqrt{12} \approx 0.9069$, and much larger than the typical jamming density for conventionally prepared packings with the same disk size distributions as the triangulated packings, $\varphi_J \simeq 0.849$~\cite{ohern_random_2002,ohern_jamming_2003}.

These processes leave the radii distribution approximately fixed relative to the radii minimized packings. We characterize this change in greater detail in end matter. 
%Figure \ref{fig:PackingWithPhi6} shows that triangulated packings with initial radii drawn from a 20\% polydispersity lognormal distribution are distributed more closely to a 26\% polydisperse log normal distribution after radii minimization. The final step of creating the perfectly triangulated packing (illustrated in Figure \ref{fig:PackingWithPhi6} typically involves changes to particles' relative radii on the order of 0.2\%. 
We subsequently produce packings at a range of pressures by increasing the packing fraction as guided by the protocol described in~\cite{charbonneau_jamming_2015}.

From each triangulated packing we produce conventionally jammed position-minimized packings with identical radii distributions by first randomizing positions and then minimizing energy with respect to only position degrees of freedom~\cite{charbonneau_universal_2012}. Conventionally jammed packings are illustrated in Figure \ref{fig:PackingWithPhi6}b.

\textit{Properties of an Ideal Packing} -- 
%\section{Properties of an Ideal Packing}
We demonstrate that our triangulated packings meet all of the requirements for an ideal jammed packing:

1) Ideal packings must be triangulated. The circle packing theorem provides a mechanism to convert any embeddable 2d graph into a disk packing with contact network equal to that graph.  Generically, there will be a very large number of disk packings that share the same contact graph and thus the configurational entropy of a packing is bounded by that of the associated contact graph.  However, for triangulated graphs there is a one-to-one mapping (unique up to trivial symmetries like rotations and mirrorings) between triangulated disk packings and triangulated graphs and thus the configurational entropy of the triangulated packing is equal to that of the underlying contact graph. As we show in the supplementary material, the specific configurational entropy of a triangulated graph goes to \textit{zero} in the thermodynamic limit and thus the specific configurational entropy of triangulated packings also must go to zero~\cite{bender_number_2002, gimenez_asymptotic_2009, chapuy_asymptotic_2011}. Finally, a non-triangulated packing will necessarily and generically have neighboring particles which are separated by small gaps. Relaxing those gaps away must then lead to a denser packing. This will be impossible for a triangulated packing, which must then be the densest packing possible with a given set of radii. By construction, the packings in this manuscript have a fully triangulated contact network at the point of jamming.

2) Ideal amorphous packings must, by definition, have no crystalline orientational or translational order.  We demonstrate this lack in triangulated packings by computing the correlation function of the orientational order parameter $\psi_6$~\cite{halperin_theory_1978}, and the finite size scaling of the of the translational order metric $\tau$~\cite{torquato_ensemble_2015}.

The local orientational order parameter, $\psi_{6,i}$, is defined for disk $i$ as
\begin{equation}
    \psi_{6,i}\equiv\frac{1}{N_i} \sum_{j=1}^{N_i} e^{6 i \theta_{ij}},
\end{equation}
where $N_i$ is the number of neighbors, $\theta_{ij}$ is the polar angle of the vector between particle $i$ and neighbor $j$~\cite{halperin_theory_1978}. The correlation function $C_6(r)$
\begin{equation}
    C_6(r) \equiv\left< \psi_6\left(\vec{r}_1\right)\psi^*_6\left(\vec{r}_2\right)\right>_{|\vec{r}_2-\vec{r}_1|=r},
\end{equation}
an average over $\psi_6$ pairs separated by distance $r$, is constant for a crystal but decays exponentially for both the conventionally jammed and triangulated amorphous systems, as shown in Figure \ref{fig:PackingWithPhi6}c. Note, $C_6$ for a crystal is always one, irrespective of distance.

The translational order metric $\tau$~\cite{torquato_ensemble_2015}, is defined in 2D space as 
\begin{equation}
    \tau\equiv\frac{1}{D^2} \int \left( g(r)-1 \right)^2 d\vec{r},
\end{equation}
where $g(r)$ is the pair correlation function and $D$ is a length scale chosen to be the average particle radius. Finite size scaling in $\tau$ indicates the degree of long range translational order in a system; Poisson distributed points have $\tau=0$ for all system sizes, amorphous disk packings maintain a constant non-zero value of $\tau$ with increasing $N$, and ordered systems show power law growth in $\tau$ with $N$. Figure \ref{fig:PackingWithPhi6}d shows $\tau$ as a function of system size for the hexagonal crystal (circles), conventionally jammed systems (tan, labelled), and triangulated packings (violet, labelled).

Taken together, these two measurements demonstrate that triangulated packings are equally amorphous as conventionally jammed packings. Note that the absence of crystalline order does not preclude other forms of amorphous order, such as those revealed by point-to-set correlations, steric correlations~\cite{fan_ideal_2024} and other many-body correlation functions.

\begin{figure}
    \centering
    \includegraphics[trim=5 20 20 0, clip, width=1\linewidth]{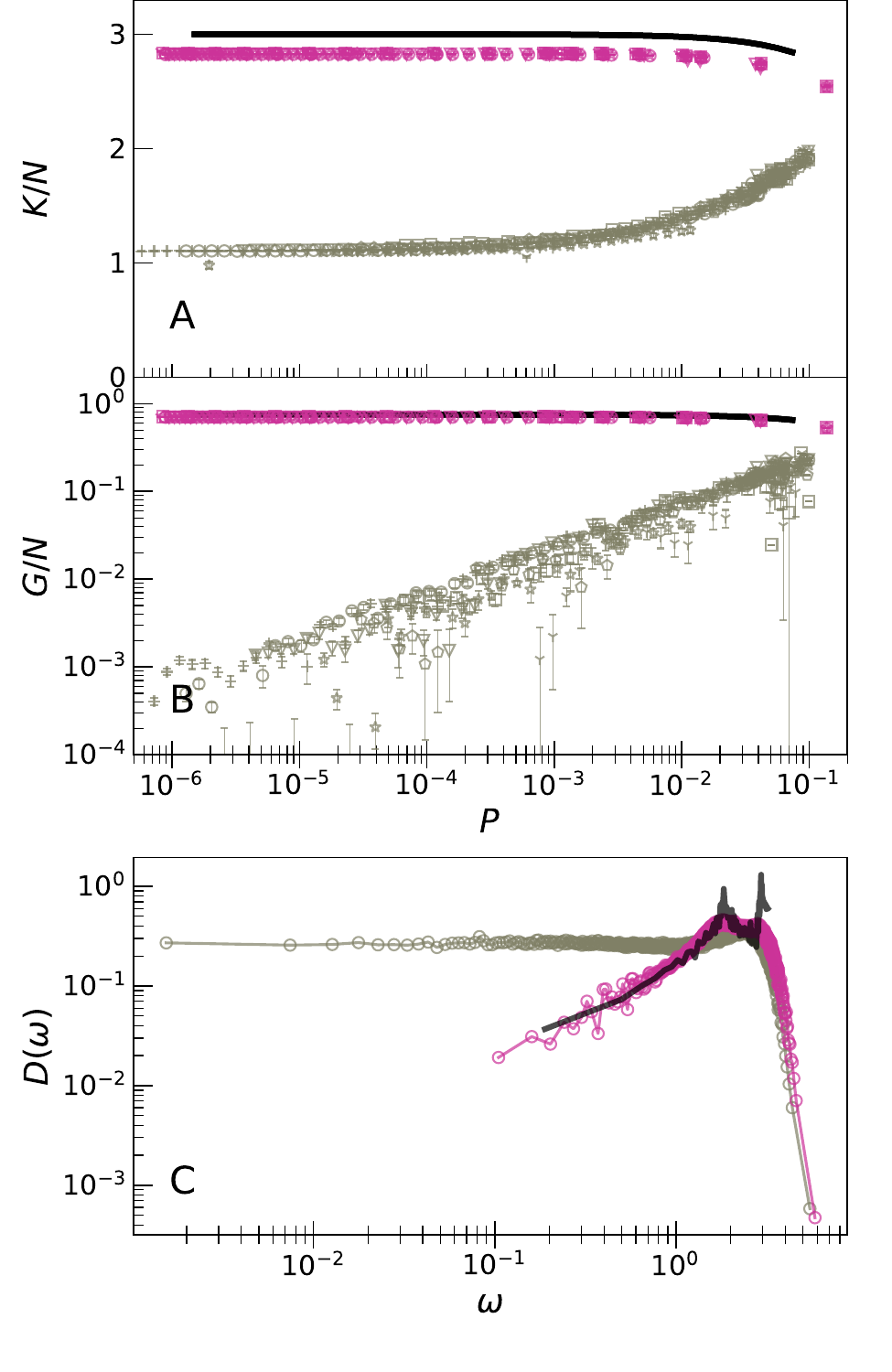}
    \caption{ a) Bulk modulus per particle $K/N$ are plotted as a function of pressure, $P$ for triangulated packings (magenta points), conventionally jammed packings (tan points), and hexagonal crystals (black line) for $N$ ranging from 64 to 4096 in powers of two (indicated by Ys, squares, pentagons, triangles, stars, +'s and o's respectively). Moduli are averaged over 10 independent packing. Error bars indicate standard error. b) Shear modulus per particle, $G/N$, is presented with the same colors and symbols. c) Density of vibrational states, $D(\omega)$, are presented with the same colors and symbols; only $N=4096$ is shown for clarity.}
    \label{fig:pressureScalingModuli}
\end{figure}

\begin{figure}
    \centering
    \includegraphics[trim=5 10 10 0, clip, width=1.0\linewidth]{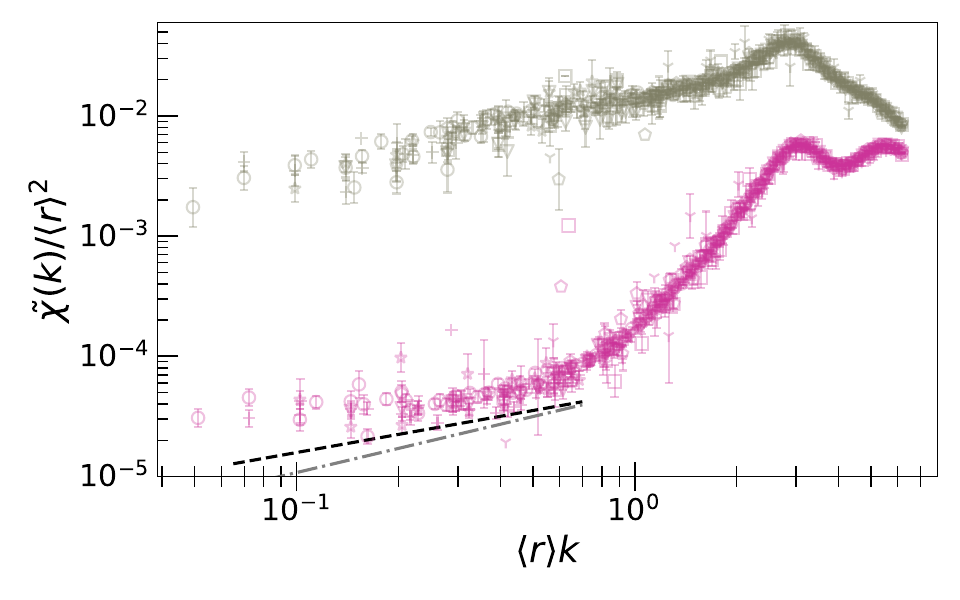}
    \caption{ Binned averages of spectral density $\tilde\chi$, scaled by $\left<r\right>$ to collapse curves from $N=64-8192$; respective markers are, in increasing powers of two, Ys, squares, pentagons, triangles, stars, +'s, o's and hexagons. Color indicates protocol; tan for position-minimized packings and magenta for triangulated packings. Dashed lines with slope of $1/2$(black) and 2/3~\cite{donev_linear_2004} (grey) are overlaid as a visual guide. Error bars indicate standard geometric error of mean for each point; points without error bars are unaveraged.}
    \label{fig:spectralDensity}
\end{figure}

\begin{figure}
    \centering
    \includegraphics[trim=0 0 0 0, clip, width=1.0\linewidth]{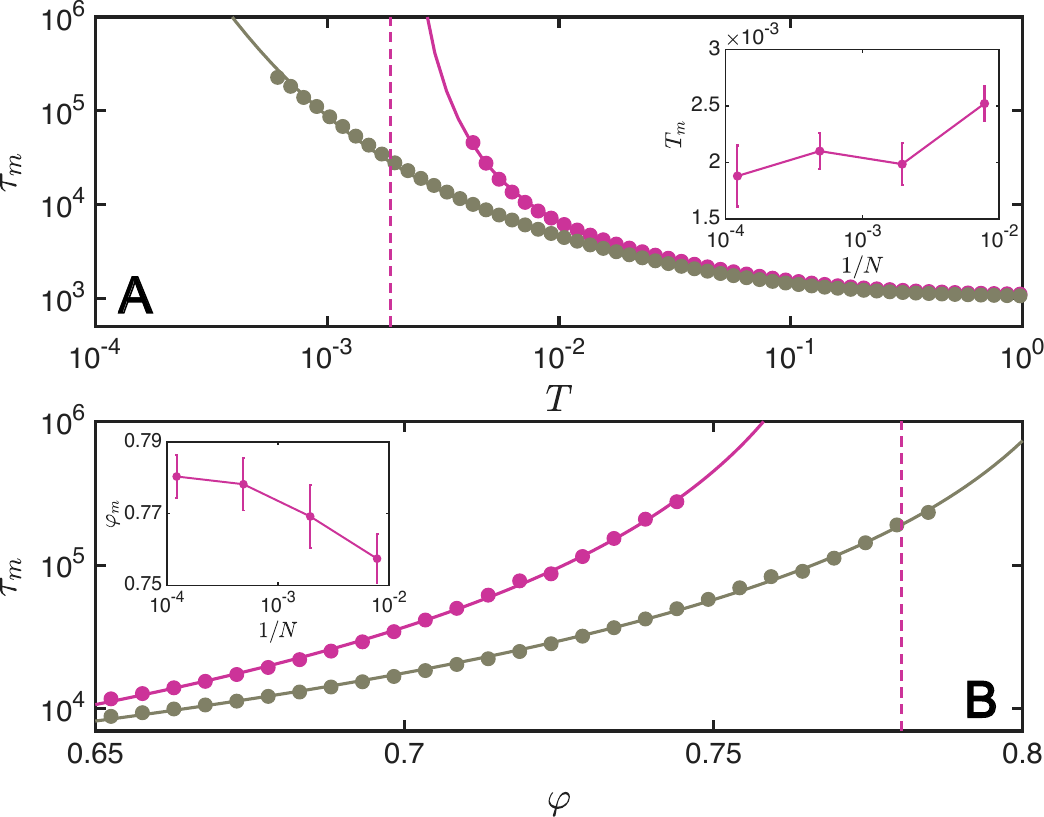}
    \caption{Measurements of the relaxation time for a sample triangulated packing (magenta) and conventionally jammed packing (tan), each with $N=8192$ using 10 thermal averages for A) thermal soft spheres and B) hard spheres. Solid line fits are to the modified VFT form of Eq.~\eqref{eq:VFT} and to the power-law diffusion form of Eq.~\eqref{eq:powerLawDiff} respectively, and dashed lines show the ideal glass melting point. Insets show a lack of statistically significant finite size scaling for $N>128$.}
    \label{fig:enter-label}
\end{figure}

% \begin{figure}
%     \centering
%     \includegraphics[trim=0 0 0 0, clip, width=1.0\linewidth]{meltingTempCompareErrorBars.pdf}
%     \caption{Measurements of the relaxation time for a sample triangulated packing (magenta) and conventionally jammed packing (tan), each with $N=8192$ using 10 thermal averages. A fit to the modified VFT form of Eq.~\eqref{eq:VFT} for the triangulated packing is overplotted and yields a divergence at $T_m = 0.00189(10)$ (indicated by a vertical dashed line), $c = 0.57(3)$, and $D=13.3(7)$ with the error bar indicating a 95$\%$ confidence interval.  The conventionally jammed packing is fit by $T_m < 10^{-4}$, $c = 0.43(1)$, and $DT_m=0.035(4)$, indicating that the divergence is still far from the lowest simulated temperature.}
%     \label{fig:enter-label}
% \end{figure}

3) Ideal packings must be mechanically ultrastable as a result of being the deepest well in the energy landscape. Ideal jammed packings must be strictly jammed, in the sense that no collective motions or strains lead to unjamming motions~\cite{donev_linear_2004, donev_jamming_2004}. By employing a linear programming algorithm~\cite{donev_linear_2004, donev_jamming_2004} that takes into consideration the one-sided contact potential, we find that all triangulated packings tested up to $N=4096$ are strictly jammed just as crystalline systems are strictly jammed.

Figure \ref{fig:pressureScalingModuli} shows the mechanical properties of conventional packings, triangulated packings, and the hexagonal crystal.  We calculate elastic moduli~\cite{dennis_dionysian_2022} and the vibrational density of states using the linear dynamical matrix of a packing. Triangulated packings are mechanically ultrastable as indicated by constant, and anomalously large, bulk and shear moduli as pressure is brought to zero, in contrast to the vanishing shear modulus at zero pressure found in conventionally jammed packings. Further, they play host to vibrational modes at vanishing pressure which are governed by the Debye, or Phononic~\cite{xu_low-frequency_2024,royall_race_2018}, scaling of crystals and ultrastable systems and lack the Boson peak~\cite{manning_random_2015,charbonneau_universal_2016} of conventionally jammed packings. Just as in a 2d crystal, there is no phonon gap in the ideal packing because there is only a single “phase” throughout the entire system.  Further, because the ideal packing is amorphous, there are no special lattice directions and thus the system is fully isotropic. In sum, the mechanical properties of triangulated packings are much more similar to those of crystals than to those of conventionally jammed packings~\cite{berthier_zero-temperature_2019}, rendering them maximally distant from the point of marginality.

4) An ideal packing must have vanishing long range density fluctuations and thus be hyperuniform~\cite{torquato_local_2003}.  If it had such fluctuations then one could cut out the lower density regions and replace them with copies of the higher density regions in order to achieve a denser packing. Further, low-entropy is a necessary condition for hyperuniformity, which is only possible for states of zero complexity~\cite{godfrey_absence_2018}.

We show that our triangulated packings are hyperuniform by directly measuring the long wavelength (and thus small wavenumber, $\vec k$) scaling in the spectral density~\cite{zachary_hyperuniform_2011}:
\begin{equation}
 \tilde{\chi}(\vec{k})\equiv\frac{\big|\sum_{j=1}^N\exp(-i\vec{k}\cdot\vec{x}_j)\tilde{m}(\vec{k};R_j)\big|^2}{V},
\end{equation}
where $V=1$ is the box volume,
%$\tilde{m}(\vec{k};R_j)$ is the Fourier transform of the indicator function for a disk of radius $R_j$ given in dimension $d$ as 
\begin{equation}
 \tilde{m}(\vec{k};R_j) \equiv \frac{2\pi R_j}{|\vec{k}|} J_1(R_j|\vec{k}|),
\end{equation}
and $J_1$ is the Bessel function of the first kind with order $1$. We take a Fourier space angular average to produce $\tilde{\chi}(k)$, shown in Figure \ref{fig:spectralDensity}. Hyperuniformity corresponds to a vanishing value of $\tilde{\chi}(k)$ in the limit of small $k\equiv|\vec{k}|$~\cite{torquato_local_2003,zachary_hyperuniform_2011}. While both conventionally jammed and triangulated packings show approximately power law behavior for small $k$
~\cite{skoge_packing_2006,tjhung_hyperuniform_2015,hexner_hyperuniformity_2015,ikeda_large-scale_2017}, the triangulated packings show a far greater degree of effective hyperuniformity~\cite{torquato_hyperuniform_2018}.

% 5) The hard sphere melting density or soft sphere melting temperature measures the thermodynamic stability of a glass and encodes information about the height of the barriers about the stable basin. In an ideal glass the melting density should be lower than that of any other glass and equal to the Kauzmann density \vbl{Check this}. To measure the melting density, we first take an ideal glass and decompress the system to a range of densities $\varphi < \varphi_J$ while keeping center positions fixed. We then apply simple constant volume Monte Carlo dynamics using standard Metropolis hard sphere sampling using the same packing fraction $\phi$ for both the conventional jammed packings and the triangulated packings. For each density, we measure the time $\tau_m$ that it takes for the system to forget its initial state (i.e., the state-overlap parameter becomes $1/e$)~\cite{charbonneau_dimensional_2022}. This then resembles a standard diffusion experiment with initial positions set by those of the ideal glass. Because of this, we can fit to a standard 

5) The melting temperature (for soft spheres) and the melting density (for hard spheres) measure the thermodynamic stability of a glass and encode information about the height of the barriers about the stable basin. In an ideal glass the melting temperature should be higher and the melting density should be lower than that of any other glass. Using standard Monte Carlo techniques, we measure the relaxation time of the thermal packing seeded by the positions of the jammed state at various temperatures (soft spheres) or deflated densities (hard spheres) to identify the point at which the relaxation time diverges. We identify this point as the melting point of each packing (see End Matter for details).

We find that the conventional packings have $T_m < 10^{-4}$ for soft spheres and $\varphi_m = 0.824(5)$ for hard spheres, while the triangulated packings have a melting temperature of $T_m=0.00189(10)$ for soft spheres and $\varphi_m = 0.780(6)$ for hard spheres. The uncertainty here indicates a 95$\%$ confidence interval, and no statistically significant finite size effects are seen for $N>128$. 

These results stand in opposition to previous work done on ultrastable (but not ideal) 2d hard sphere glasses which can be read to suggest $T_m = 0$ for finite size systems~\cite{berthier_zero-temperature_2019, santen_absence_2000}.  This distinction is likely due to the difference between truly ideal systems and those which are merely ultrastable.

Of course, the Mermin-Wagner-Hohenberg (MWH) theorem~\cite{hohenberg_existence_1967,mermin_absence_1966} would seemingly require that the melting temperature in 2d should converge to zero in the thermodynamic limit.  We see no evidence of such convergence. MWH assumes continuous symmetries for which the length scale diverges. This is why MWH is seemingly violated in, for example, the KTHNY scenario~\cite{sampedro_ruiz_melting_2019} where this scale is extremely large, yet finite. Further, recent work has questioned the applicability of MWH to amorphous phase transitions~\cite{vivek_long-wavelength_2017,shiba_unveiling_2016,illing_merminwagner_2017}.

\textit{Conclusion} -- The nature of the glass transition remains a mystery, and the question of whether an ideal configuration exists lies at the heart of this mystery~\cite{berthier_theoretical_2011}. The constructive scheme presented herein demonstrates not only that ideal packings exist but that they have superlative structural, mechanical, and thermal properties.  This represents a leap forward in understanding the complex energy landscapes underlying disordered systems and places the existence of a thermodynamic glass phase on firm footing. By working backwards from the ideal system through the introduction of defects we hope to fully explore the glassy landscape of two-dimensional amorphous systems. Ideal packings are also interesting from a practical perspective. Such packings are fully amorphous while presenting the bulk material and thermal properties of crystalline materials: ultrastability, hyperuniformity, and a high melting temperatures; all properties that are desirable from an industrial perspective. However, novel approaches will be necessary to create such packings in practice, as they are not accessible through common thermal or mechanical processes.  To create such systems in practice a physical implementation of our algorithm would have to be developed.

\textit{Acknowledgements} -- We thank Francesco Arceri, Ludovic Berthier, Varda Hagh, Sascha Hilgenfeldt, and Ken Stephenson for valuable discussions.  This work benefited from access to the University of Oregon high performance computer, Talapas.  This work was supported by the Simons Foundation No. 454939 (V.B.L., R.C.D, and E.I.C.).

Data are available upon request. CirclePack is available at circlepack.com.

\bibliography{PRLSubmission/idealGlass}
\clearpage
\onecolumngrid
{\centering {\huge \textbf{End Matter} \par}}
\setcounter{section}{0}
\setcounter{figure}{0}
\renewcommand{\thefigure}{A\arabic{figure}}
\bigskip

\twocolumngrid

\section{Diameter Distribution}

\begin{figure}[h]
\centering
\includegraphics[trim={12 16 8 2}, clip, width=.98\linewidth]{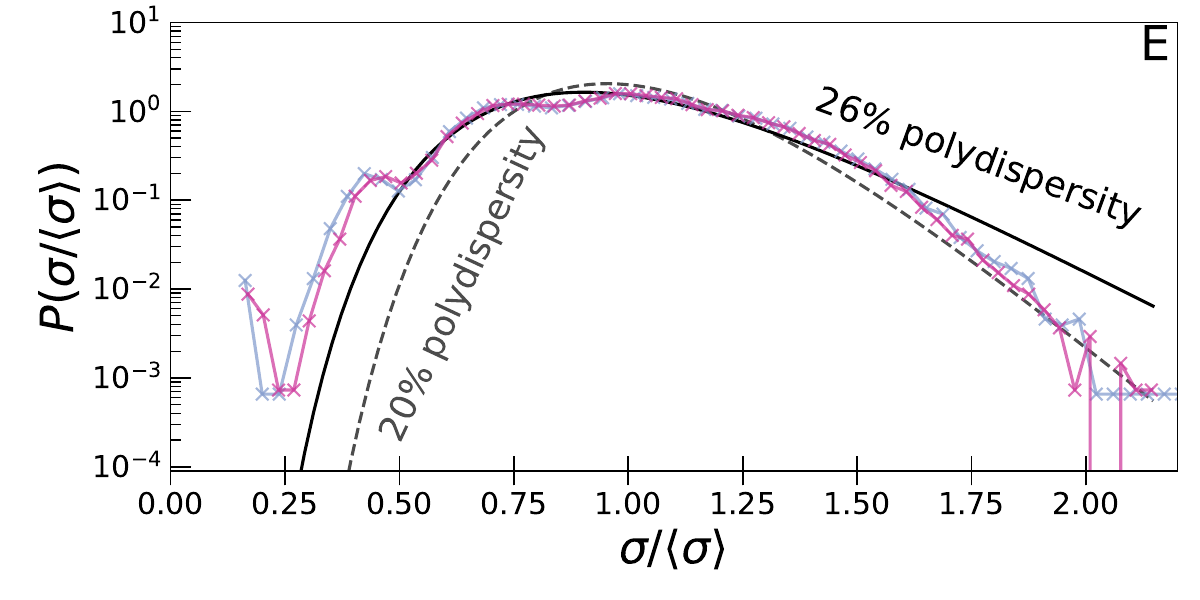}
    \caption{The probability distribution of disk diameters, $P(\sigma)/\langle \sigma \rangle)$ averaged over 10 packings of 4096 disks.  The starting distribution is at 20\% polydispersity (gray dashed line), the distribution after radii-minimization is plotted as gray x's, the distribution after triangulation is plotted as magenta triangles, and for comparison a 26\% polydispersity distribution is shown in black.}
    \label{fig:radius distribution}
\end{figure}
Figure \ref{fig:radius distribution} shows that triangulated packings with initial radii drawn from a 20\% polydispersity lognormal distribution are distributed more closely to a 26\% polydisperse log normal distribution after radii minimization. The final step of creating the perfectly triangulated packing (illustrated in Figure \ref{fig:PackingWithPhi6} typically involves changes to particles' relative radii on the order of 0.2\%. Note that that radius minimized packings are already quite close to a fully triangulated state, and thus require only minimal changes in radii.

\section{Measuring the melting point}

To measure the melting temperature, we apply simple constant volume Monte Carlo dynamics at a range of temperatures $T$ using standard Metropolis sampling with the Boltzmann weight $\exp(-\Delta U/T)$~\cite{frenkel_understanding_2002} using the same packing fraction $\phi$ for both the conventional jammed packings and the triangulated packings. For each temperature, we measure the time $\tau_m$ that it takes for the system to forget its initial state (i.e., the state-overlap parameter becomes $1/e$)~\cite{charbonneau_dimensional_2022}. The melting temperature $T_m$ is then extracted from a fit to a modified Vogel-Fulcher-Tammann (VFT) form~\cite{garcia-colin_theoretical_1989}. 
\begin{equation}
\tau_m \equiv\tau_0\exp\bigg[\bigg(\frac{DT_m}{T-T_m}\bigg)^c\bigg]
\label{eq:VFT}
\end{equation}
where $D$ is a dimensionless material dependent constant. The standard VFT equation with $c=1$ fits our data poorly, in large part due to the large polydispersity, which generically leads to a wide variety of minima, and thus dynamical heterogeneities. We thus take inspiration from other models with a high degree of dynamical heterogeneity which suggest a stretched exponential form in the viscosity and relaxation times~\cite{chakraborty_glasses_2022, ritort_glassy_2003}, and thus a variable $c$. This exponent then relates to the ductility of the glass, with higher values of $c$ indicating stronger glasses~\cite{debenedetti_supercooled_2001}.

%We perform $10$ thermal averages on each sample, average the values of $\tau_\alpha$, and perform a fit to determine $T_m$. 
We find that the conventional packings have $T_m < 10^{-4}$, while the triangulated packings have a melting temperature at least an order of magnitude larger, $T_m=0.00189(10)$ with the error bar indicating a 95$\%$ confidence interval. Here, the melting temperature remains constant and non-zero to within the reported fitting error for system sizes $N>128$, indicating a lack of finite-size scaling. The presence of a truly non-zero melting temperature in a 2d system suggests a violation of the Mermin-Wagner-Hohenberg (MWH) theorem, which forbids spontaneous symmetry breaking. There are at least two caveats to consider: 1), it is possible that an amorphous glass breaks the continuous symetry of the liquid and thus does not require a violation of MHW~\cite{vivek_long-wavelength_2017,shiba_unveiling_2016,illing_merminwagner_2017}, 2) the KTHNY scenario~\cite{sampedro_ruiz_melting_2019} offers the possibility of a transition between quasi-long-ranged and long-ranged ordering which also would not violate MWH.

% We find that the conventional packings have $T_m < 10^{-4}$, while the triangulated packings have a melting temperature at least an order of magnitude larger, $T_m=0.00189(10)$ with the error bar indicating a 95\% confidence interval. While in principle this should correspond to a Kauzmann temperature, its existence in 2-dimensions has been debated~\cite{berthier_zero-temperature_2019,santen_absence_2000}. \vbl{Here, the melting temperature remains constant and non-zero to within the reported fitting error for system sizes $N>128$, indicating a lack of finite-size scaling}. The presence of a truly non-zero melting temperature in a 2d system \vbl{suggests} a violation of the Mermin-Wagner-Hohenberg theorem, which forbids spontaneous symmetry breaking in the thermodynamic limit.  However, as discussed in Vivek et al~\cite{vivek_long-wavelength_2017} and several others~\cite{shiba_unveiling_2016,illing_merminwagner_2017} it is not a priori clear that an amorphous glass breaks the continuous symmetry of the liquid and thus it may be possible for this transition to exist at finite temperature, even in the thermodynamic limit~\cite{vivek_long-wavelength_2017}.

The process of determining the melting density follows a similar routine by using constant volume Monte Carlo dynamics at different packing fractions $\varphi$ with hard sphere Metropolis sampling starting from the seeded ideal (or non-ideal) glass positions. Here, however, the relaxation follows a simple power law scaling,
\begin{equation}
\tau_m = D(\varphi_m-\varphi)^{-\gamma}
\label{eq:powerLawDiff}
\end{equation}
where for the ideal glass at $N=8192$, we find $D=57(12)$, $\varphi_m=0.780(6)$, and $\gamma=2.6(3)$, and for the non-ideal glass, we find $D=150(30)$, $\varphi_m=0.824(5)$, and $\gamma=2.28(15)$. Curiously, we find that this seeded diffusion process has a critical exponent $\gamma$ which is consistent with the mean-field glass value of $\gamma=2.33786$~\cite{kurchan_exact_2013} unlike the monodisperse counterparts~\cite{charbonneau_dimensional_2022} which undershoot this value. As with the melting temperature, we observe no statistically significant finite size scaling for $N>128$, indicating that the thermodynamic melting density is $\varphi_m=0.780(6)$.

\end{document}

\title{The Ideal Glass and the Ideal Disk Packing in Two Dimensions: Appendix}
\author{Viola M. Bolton-Lum, R. Cameron Dennis, Peter K. Morse and Eric I. Corwin}

%% Comment out or remove this line before generating final copy for submission; this will also remove the warning re: "Consecutive odd pages found".
%\instructionspage  

\maketitle

In this appendix we develop an argument that a maximal graph with periodic boundary conditions on a torus (genus 1) has zero specific entropy in support of our manuscript. Enumerating circle packings is a much more difficult problem than enumerating graphs, but as we will see, this is an appropriate proxy for predicting the specific entropy of circle packings. The core of the argument is that in the limit of the number of nodes going to infinity, the proportion of graphs that are maximal goes to zero. We will outline why this is true and then demonstrate how it relates to packings of circles in periodic boundary conditions.

%begin copy part 3
We limit our argument to graphs which are 2-connected because this is what is available in the literature.

Put in simple terms, a $k$-connected graph is a graph where a path exists between every vertex even after removing any set of $k-1$ vertices and their edges. From this definition, any $k$-connected graph is 2-connected for $k>=2.$ It is wise, in this analysis, to omit graphs which are not 2-connected as these will never represent jammed configurations. A graph which is not 2-connected has at least one vertex for which multiple disconnected clusters will form upon its removal. Such a vertex is termed an articulation point and the independent clusters can pivot about this point. Such a graph is floppy and we should not consider it among the set of jammed packings. However, it is worth noting that the inverse is not necessarily true: a specific $k$-connected graph for $k>=2$ does not necessarily have corresponding jammed packings with the underlying graph structure.
%end copy part 3

%begin copy part 1
We further restrict our analysis to packings of hard disks with at least $3$ contacts per disk. It is easy to see that this is required for local rigidity. However, it is worth noting again that having three contacts per particle is not a sufficient condition for jamming. To better reduce our search space, we consider the Maxwell-Calladine index theorem which states that a two dimensional elastic network with $n$ nodes (vertices), $m$ contacts (edges), $n_\textrm{floppy}$ floppy modes, and $S$ states of self stress has the property that
\begin{align}
    n_\textrm{floppy}-S=2n-m.
    \label{eq:indexTheorem}
\end{align}

If we define the average number of edges per node, $\mu\equiv m / n$ for a network and note that we require $1$ state of self stress (from the fact that our boundaries are fixed) and $2$ trivial floppy modes, we find that Eqn. \ref{eq:indexTheorem} sets a lower bound on $\mu$ for a jammed packing as
\begin{align}
 n_\textrm{floppy}-S=2n -m \implies 2 - 1 = 2n - m \implies \mu_\textrm{min} = 2 - 1/n.
\end{align}
Note that this ignores the possibility of non-linearly rigid jammed packings, which can generally be ignored for large system sizes~\cite{holmes-cerfon_enumerating_2016}.

If we consider triangulated packings, we have the condition that every face of our network must have exactly 3 edges. If we factor in the double-counting, then for a maximal graph, $3F=2m.$ The Euler characteristic for a toroidal graph is $\chi=n- m + F = 2-2g=0$, where $F$ is the number of faces and $g=1$ is the genus of a torus. Using $3F=2m$, the relation between the number of faces and edges for a maximal graph, the maximal value of $\mu$ is
\begin{align}
    \mu_{\textrm{max}}=3.
\end{align}
Note again that there exist packings of hard disks that are not jammed and yet may satisfy the property $\mu>=\mu_{\textrm{min}}.$

%end copy part 1

Now that we have defined our search space, we begin our derivation with equation 4 from Chapuy et. al.\cite{chapuy_asymptotic_2011}. The number of graphs, $b_{n,m}^{(g)}$, with $n$ nodes, $m$ edges, and of genus $g$ has the following asymptotic form
\begin{align}
b_{n,m}^{(g)}\sim d_\mu^{(g)}n^{5g/2-4}\left(\delta_{\mu}\right)^n n!\label{eq:torusB}
\end{align}
with $n$ vertices and $m = \lfloor \mu n\rfloor$ edges, where $d_\mu^{(g)}$ is a prefactor that depends on surface genus and $\delta_\mu$ is the base of the exponential growth with $n$ and does not depend on genus.

%begin copy part 2
This work by Chapuy et al. is an extension of the work by Bender et al. \cite{bender_number_2002} which is valid for planar graphs (genus $g=0$). In Bender et al., the authors have the following theorem for planar graphs (in the notation of~\cite{chapuy_asymptotic_2011}):
{\it For $m_0/n\in J,$ there is a unique $t\in(0,1),$ such that $\mu(t)=m_0/n,$ and }
\begin{align}
b_{n,m}^{(0)}=\frac{3x_0(t)^2 y_0(t) D_3(t)n!}{8\sqrt{2}\pi\left(1+y_0(t)\right)\sigma(t) n^3 m}x_0(t)^{-n}y_0(t)^{-m}\left(\exp\left\{-\frac{\left(m - m_0\right)^2}{2n\sigma^2(t)}\right\} + o(1)\right),\label{eq:planarB}    
\end{align}
{\it uniformly as $n\to \infty$ and $m_0/n\in J.$} In this expression $x_0, y_0,$ and $\sigma$ are all pure functions of a variable $t$. We work to massage the equation into the form given by~\ref{eq:torusB}. Without loss of generality, the authors of Bender et al.~\cite{bender_number_2002} choose $t$ such that $t=t(1)\approx 0.62637$ and $y_0(t)=1.$ This value of $t$ gives
\begin{align}\sigma=0.618431,~x_0=0.03819,~\textrm{and~} \mu_0=2.2629.\label{equation:1}
\end{align}
If we let $\mu\equiv m/n,$ then
\begin{align}
b_{n,m}^{(0)}=\frac{3x_0^2 D_3n!}{16\sqrt{2}\pi\sigma n^4 \mu}x_0^{-n}\left(\exp\left\{-\frac{n\left(\mu - \mu_0\right)^2}{2\sigma^2}\right\} + o(1)\right).
\end{align}
This can be rewritten asymptotically as
\begin{align}
b_{n,m}^{(0)}\sim d_{\mu}^{(0)}n^{-4}\left(\delta_{\mu}\right)^nn!
\end{align}
by gathering the prefactor into $d_\mu^{(g)}$ and defining the base of the exponential as
\begin{align}
    \delta_{\mu}\equiv \left(\frac{1}{x_0}\exp\left\{\frac{\left(\mu-\mu_0\right)^2}{2\sigma^2}\right\}\right).
\end{align}
%end copy part 2

We now consider the ratio of $b_{n,\mu_2 n}^{(1)}$ and $b_{n,\mu_1 n}^{(1)}$ which we define as
\begin{align}
    R_{\mu_1,\mu_2}&\equiv \frac{b_{n,\mu_2 n}^{(1)}}{b_{n,\mu_1 n}^{(1)}} =\left(\frac{d_{\mu_2}^{(1)}}{d_{\mu_1}^{(1)}}\right)\left(\frac{\delta_{\mu_2}}{\delta_{\mu_1}}\right)^n\\
&=C_{1,2}\exp\left\{-\frac{1}{2\sigma^2}\left(\mu_2-\mu_1\right)\left(\mu_2+\mu_1-2\mu_0\right)\right\}^n,
\end{align}
where $C_{1,2}$ is the ratio of prefactors and relatively unimportant.  We compute the ratio of the number of maximal toroidal graphs to the number of toroidal graphs for any value of $\mu_1 < 3.$ Let $\mu_2=3$  represent the case of maximal toroidal graph. We see that in the limit as $n\to \infty,$ this ratio tends to 0 when 
\begin{align}
    3>\mu_1>2\mu_0-3\approx 1.5258.
\end{align}
Since these equations do not apply to graphs for which $\mu<2-1/n,$ we have shown that the ratio of maximal toroidal graphs to any other type of 2-connected toroidal graph tends to 0 in the thermodynamic limit.

\section*{Zero Specific Entropy}
From the circle packing theorems~\cite{stephenson_introduction_2005}, we know that maximal graphs correspond uniquely to circle packings. However, this is not true for graphs with lower coordination. In fact, from the circle packing theorems, we know that there is at least one circle packing for every graph. Given these two facts, we know that $R_{\mu,3}$ is an upper bound for the ratio of the number of triangulated circle packings to circle packings with a fixed $\mu$ value where $3>\mu\geq 2.$ Since this upper bound tends to $0$ in the thermodynamic limit, the specific entropy of maximal circle packings is $0$ in the thermodynamic limit. We note that there are many pairs $\mu_1 \neq \mu_2$ for which $R_{\mu_1, \mu_2}$ will also tend to zero.  However, $\mu_2=3$ is the only value for which this ratio will \textit{always} tend to zero.

\section*{Configurational Entropy}
It is true that the specific entropy of maximal graphs goes to $0$ in the thermodynamic limit, but one may also be interested in the behavior of the configurational entropy. 

For this, we use equation 4 in reference \cite{chapuy_asymptotic_2011} which is the formula for the total number of 2-connected graphs on a torus (for all possible values of $\mu$),
\begin{align}
    b_n^{(1)}\sim d^{(1)}n^{-1}\delta^n n!
\end{align}
and reference \cite{bender_number_2002} tells us that
\begin{align}
    \delta = 1 / x_0.
\end{align}

We define a new ratio $Q_{\mu}$ as the fraction of \textit{all} graphs with $n$ nodes that happen to have a particular value of $\mu$ to be
\begin{align}
    Q_{\mu}&\equiv\frac{b_{n,\mu n}^{(1)}}{b_n^{(1)}}\\
    &= \frac{d_{\mu}^{(1)}}{d^{(1)}}n^{-1/2}\exp\left\{-\frac{1}{2\sigma^2}\left(\mu-\mu_0\right)^2\right\}^n.
\end{align}
From this it is clear that any $\mu$ other than $\mu=\mu_0$ is rare in the thermodynamic limit.

Next we define the asymptotic probability of finding a graph with a given $\mu$ and $n$
\begin{align}
    P(\mu,n)\sim P_0 n^{-1/2}\exp\left\{-\frac{1}{2\sigma^2}\left(\mu - \mu_0\right)^2\right\}^n.
\end{align}
From this, we define the asymptotic configurational entropy of our system,
\begin{align}
    S(\mu, n) \sim -k_B P(\mu, n)\ln P(\mu,n).
\end{align}
After substituting,
\begin{align}
    S(\mu,n)&\sim  -k_B P_0\exp\left\{-\frac{1}{2\sigma^2}\left(\mu-\mu_0\right)^2\right\}^n\left[\ln P_0 n^{-1/2}-\frac{1}{2}n^{-1/2}\ln n - \frac{1}{2\sigma^2}n^{1/2}\left(\mu - \mu_0\right)^2\right].
\end{align}

We wish to know the value of $\mu$ which gives the lowest configurational entropy. We first consider the boundaries by substituting the numerical values found earlier in~\eqref{equation:1} :
\begin{align}
    S(3,n)\sim k_BP_0\left(0.491475\right)^n\left[-\ln P_0 n^{-1/2}+\frac{1}{2}n^{-1/2}\ln n +0.710344n^{1/2}\right]
\end{align}
and
\begin{align}
    S(2,n)\sim k_BP_0\left(0.913619\right)^n\left[-\ln P_0 n^{-1/2}+\frac{1}{2}n^{-1/2}\ln n +0.0903418n^{1/2}\right]
\end{align}

For large $n,$ $S(3,n)$ is lower than $S(2,n)$ therefore, $S(2,n)$ is not a minimum in the limit of large $n.$ We check if there are any critical points which may correspond to minima. By differentiating $S,$ this condition is satisfied when
\begin{align}
\exp\left\{-\frac{1}{2\sigma^2}\left(\mu-\mu_0\right)^2\right\}^n\left(\mu-\mu_0\right)\left(-2\sigma^2+n(\mu-\mu_0)^2+\sigma^2\ln n-2\sigma^2\ln P_0\right)=0
\end{align}
The exponential term can never be $0$ so we have critical points at
$\mu=\mu_0$
and
$\mu=\mu_0\pm\sqrt{\frac{\sigma^2}{n}\left[2-\ln \left(\frac{n}{P_0^2}\right)\right]}.$

We define this last solution as $\mu_{1\pm}.$ The critical point for $\mu_0$ is
\begin{align}
    S(\mu_0,n)\sim  k_B P_0\frac{1}{2\sqrt{n}} \ln \frac{n}{P_0^2}.
\end{align}
This decays very slowly with $n$ so we can easily rule this out as a minimum for large $n.$ Finally, we check
\begin{align}
    S(\mu_{1\pm},n)\sim \frac{ k_B}{e}\end{align}
This entropy is constant with $n$ and will therefore never be the lowest entropy state.

From this, we can conclude that the configurational entropy of $\mu=3$ approaches $0$ for 2-connected toroidal graphs faster than any other value. For sufficiently large $n,$ it will be the lowest entropy state. Similarly from before, if we set $\mu=3,$ we see that this is a lower bound for the configurational entropy of maximal circle packings.

\bibliography{idealGlass}